\def \cm{~\rm{cm}}
\def \s{~\rm{s}}
\def \km{~\rm{km}}

\def \K{~\rm{K}}

\def \erg{~\rm{erg}}
\def \yrs{~\rm{yrs}}
\def \yr{~\rm{yr}}

\def \kev{~\rm{keV}}
\documentclass[12pt,preprint]{aastex}
\usepackage{graphics,epsf}
\usepackage{amsmath}                
\usepackage{amsfonts}               
\usepackage{amssymb}                
\usepackage{epsfig}            

\begin{document}

\title{X-RAY EMISSION FROM JET-WIND INTERACTION IN PLANETARY NEBULAE}

\author{Muhammad Akashi, Yohai Meiron, and Noam Soker}
\affil{Department of Physics, Technion$-$Israel Institute of
Technology, Haifa 32000, Israel
akashi@physics.technion.ac.il; ym@physics.technion.ac.il; soker@physics.technion.ac.il}

\begin{abstract}
We conduct 2D numerical simulations of jets expanding into the slow wind of
asymptotic giant branch stars. We show that the post-shock jets' material can explain
the observed extended X-ray emission from some planetary nebulae (PNs).
Such jets are thought to shape many PNs, and therefore it is expected that this
process will contribute to the X-ray emission from some PNs.
In other PNs (not simulated in this work) the source of the
extended X-ray emission is the shocked spherical wind blown by the central star.
In a small fraction of PNs both sources might contribute, and a two-temperatures gas
will fit better the X-ray properties than a one-temperature gas.
A spacial separation between these two components is expected.

\end{abstract}


\section{INTRODUCTION}

The physical mechanisms responsible for the shaping of planetary nebulae (PNs)
have been the focus of many studies in the last two decades (Balick \& Frank 2002 and
references therein).
One of the popular shaping processes involves jets launched by the
central star or its companion (e.g., Morris 1987; Soker 1990;
Sahai \& Trauger 1998; Soker \& Rappaport 2000; see Soker \& Bisker 2006 for many
more references and a discussion of the development of the jet shaping models).
The shaping jets are blown during the late AGB phase or early post-AGB phase.
As jets are launched at speeds of $\sim 300-3000 \km \s^{-1}$, they become a
source of extended X-ray emission when they are shocked (Soker \& Kastner 2003).
However, in many, and possibly majority of, cases the extended X-ray emission comes
from the shocked fast wind blown by the central star during post-asymptotic giant
branch (AGB) phase, rather than from shocked jets.
Both the jets (or CFW for collimated fast wind) and the central star wind
play significant role in shaping PNs (e.g., Balick \& Frank 2002,
and references therein). Therefore, the X-ray band can be a key to explore
the nature of shaping processes in PNs.

In recent years many PNs were found to be sources of extended X-ray emission.
PNs were detected mainly by the {\it Chandra} X-ray Observatory (CXO),
e.g., BD~+30$^\circ$3639 (Kastner et al.\ 2000; Arnaud et al.\ 1996
detected X-rays in this PN already with ASCA),
NGC 7027, (Kastner, et al.\ 2001), NGC 6543 (Chu et al.\ 2001),
Henize 3-1475 (Sahai et al. 2003), Menzel 3 (Kastner et al.\
2003),  and by the {\it XMM-Newton} X-ray Telescope, e.g., NGC
7009 (Guerrero et al.\ 2002), NGC 2392 (Guerrero et al.\ 2005),
and NGC 7026 (Gruendl et al. 2004).
This sample, and the expected future observations, motivate us to study
the X-ray emission properties of shocked jets.

One of the key unsolved problems related to X-ray emission form PNs is
the low temperature of the X-ray emitting gas.
The velocities of the winds observed during the PN phase are in most cases
$v_f > 1000 \km \s^{-1}$, which are shocked to temperatures of $T \ga 10^7 \K$.
However, the gas responsible for most of the X-ray emission has a temperature
of $\sim 1-3 \times 10^{6} \K$ (Kastner et al. 2008).
There are three possible answers to this puzzle (see Akashi et al. 2007).
(1) The hot ($T> 10^7 \K$) post-shock gas is cooled via heat conduction to the cooler
nebular gas (Soker 1994; Zhekov \& Perinotto 1996; Steffen et al. 2005), or mixing
with the cooler gas (Chu et al. 1997) enhanced by instabilities
(Stute \& Sahai 2006).
(2) The X-ray emitting gas comes mainly from a slower moderate-velocity
wind of $v_f \sim 500 \km \s^{-1}$ blown by the central star during
the post-AGB phase (Soker \& Kastner 2003; Akashi et al. 2006, 2007).
(3) The X-ray emitting gas comes mainly from two opposite jets (or CFW, Soker \& Kastner 2003),
expanding with velocities of $\sim 300-700 \km \s^{-1}$.

In the present paper we explore the third possibility by 2D hydrodynamical numerical
simulations.
We neglect heat conduction, mixing, and spherical wind blown by the central star.
This does not mean that heat conduction, mixing, or spherical wind don't
play any role. We only try to find whether in some cases
the extended X-ray emission might be fully or partially attributed to jets.

\section{NUMERICAL SIMULATIONS AND PARAMETERS}

The simulations were performed using Virginia Hydrodynamics-I
(VH-1), a high resolution multidimensional astrophysical
hydrodynamics code developed by John Blondin and co-workers
(Blondin et al. 1990; Stevens et al., 1992; Blondin 1994). We have
added radiative cooling to the code at all temperatures $T > 100 \K$.
Radiative cooling is carefully treated near contact
discontinuities, to prevent large temperature gradients
from causing unphysical results. The cooling function $\Lambda (T)$
(for solar abundances) that we used was taken
from Sutherland \& Dopita (1993; their table 6).

We simulate axisymmetrical morphologies. This allows us to use
axisymmetrical grid, and to simulate one quarter of the meridional
plane. There are 208 grid points in the azimuthal ($\theta$)
direction of this one quarter and 208 grid points in the radial direction.
The radial size of the grid points increases with radius.
In these simulations the grid extends from $10^{15} \cm$ to $4 \times 10^{17} \cm$.

Before the CFW (jet) is launched at $t=0$ the grid is filled with slow
wind having a speed of $v_1$ and mass loss rate of $\dot M_1$.
We launch a collimated fast
wind from the first 20 zones attached to the inner boundary of the grid.
The CFW is uniformly ejected within an angle (half opening angle)
$\alpha$ ($ 0 \le \theta \le \alpha$). For numerical reasons a weak slow wind is
injected in the sector $ \alpha < \theta \le 90^\circ$.
Here $\theta =0$ is along the symmetry axis (vertical in the figures).

\section{THE FLOW STRUCTURE}
\label{general}

\subsection{The different numerical runs}

Different PNs have different shapes of the X-ray emitting regions.
We emphasize that here we limit ourself to consider CFW (jets) expanding
into a spherical medium, and we do not consider other effects, such as
equatorial mass ejection and spherical fast wind blown by the central star.

The first question we address is the influence of the jet's opening angle on the
X-ray properties.
We run several cases of CFW with a velocity of $v_2=500 \km \s^{-1}$
and mass loss rate into one jet of $\dot M_2 = 2 \times 10^{-7} M_\odot \yr^{-1}$.
The CFW expands into a slow dense spherical wind that in most runs propagates at
a speed of  $v_1=10 \km \s^{-1}$ and has a mass loss rate (into all directions) of
$\dot M_1 = 10^{-5} M_\odot \yr^{-1}$ in most runs.
We simulate the CFW-Wind interaction for many values of the half opening angle,
from $10^\circ$ to $90^\circ$ (spherical wind).
We take the CFW velocity to be constant in some runs, while in other runs at time  $t=\tau_s$,
the mass loss rate starts to decrease according to the following relation
\begin{equation}
\frac{\dot M_2}{\dot M_{20}} =
\begin{cases}
 1  & \qquad  0 \le t < \tau_s
 \\
 (1- (t-\tau_s)\tau_d^{-1})^2   & \qquad  \tau_s  \le t ,
\end{cases}
\label{power1}
\end{equation}
where $\tau _d$ is a constant. In the runs where we gradually shut
down the fast wind according to equation (\ref {power1}) we take $\tau_s=1000 \yr$
and $\tau_d=1100 \yr$. We choose it because after $2000 \yrs$ the
jet mass loss rate becomes $\sim 10^{-9} M_\odot \yr^{-1}$.

We simulate the PNs for total time of $2000 \yrs$, as most X-ray bright PNs have
comparable age or are younger.
We set the grid size accordingly at $4 \times 10^{17} \cm$.
We end the simulations when the X-ray bubble exists the grid.
The different runs are summarized in Table 1.
In models Y1-Y6 we gradually shut down the CFW at $t=\tau_s=1000 \yr$
and according to equation (\ref {power1}).
\begin{table}

Table 1: Cases Calculated

\bigskip
\begin{tabular}{|l|c|c|c|c|c|c|c|}
\hline
Run & $\dot M_1$ & $\dot M_2$   & $v_2$ & $\alpha$ & Shut down the jet \\
&$M_\odot \yr^{-1}$ & $M_\odot \yr^{-1}$ &$\km \s^{-1}$ &  \\
\hline
Y1 &  $10^{-5}$ &$2\times 10^{-7}$ &$500$ & $30^{\circ}$& yes\\
\hline
Y2 &  $10^{-5}$ &$2\times 10^{-7}$ &$500$ & $40^{\circ}$ &yes\\
\hline
Y3 & $10^{-5}$ &$2\times 10^{-7}$ &$500$ & $50^{\circ}$& yes\\
\hline
Y4 & $10^{-5}$ &$2\times 10^{-7}$ &$500$ & $60^{\circ}$& yes\\
\hline
Y5 & $10^{-5}$ &$2\times 10^{-7}$ &$500$ & $75^{\circ}$ &yes\\
\hline
Y6 & $10^{-5}$ &$2\times 10^{-7}$ &$500$ & $90^{\circ}$& yes\\
\hline
Y7 & $10^{-5}$ &$1\times 10^{-7}$ &$500$ & $50^{\circ}$ & No\\
\hline
Y8 & $10^{-5}$ &$2\times 10^{-7}$ &$500$ & $50^{\circ}$& No\\
\hline
Y9 & $10^{-5}$ &$4\times 10^{-7}$ &$500$ & $50^{\circ}$ & No\\
\hline
Y11 & $10^{-5}$ &$2\times 10^{-7}$ &$1000$ & $50^{\circ}$& No\\
\hline
Y12 & $10^{-5}$ &$2\times 10^{-7}$ &$2000$ & $50^{\circ}$ & No\\
\hline
Y13 & $10^{-5}$ &$2\times 10^{-7}$ &$866$ & $50^{\circ}$ & No\\
\hline
Y14 & $10^{-5}$ &$6.3\times 10^{-7}$ &$500$ & $50^{\circ}$ & No\\
\hline
Y15 & $10^{-5}$ &$6.3\times 10^{-7}$ &$866$ & $50^{\circ}$ & No\\
\hline
Y16 & $3\times10^{-5}$ &$2\times 10^{-7}$ &$500$ & $50^{\circ}$ & No\\
\hline
Y17 & $3.33\times 10^{-6}$ &$2\times 10^{-7}$ &$500$ & $50^{\circ}$ & No\\
\hline
\end{tabular}

\footnotesize
\bigskip

Notes: (1) For all runs $v_1=10 \km \s^{-1}$.\normalsize
(2) 'Yes' means that we shut down the jet gradually according to
(eq. \ref{power1}) and 'No' that we do not.\normalsize
\end{table}

\subsection{Results}

As mentioned in section 1, one of the main questions regarding X-ray emission from PNs
is wether the X-ray comes from a CFW blown by a companion or/and comes from spherical fast
wind blown by the post-AGB central star.
Akashi et al. (2006, 2007) have found that the observed X-ray emission of PNs can be
accounted for by shocked wind segments that were expelled during the pre-PN and early PN
phases, if the fast wind speed is moderate, $\sim 400-600 \km \s^{-1}$,
and the mass loss rate is $ {\rm few}\times 10^{-7} M_\odot \yr^{-1}$.
Here we investigate the X-ray emission evolution for CFW-Wind, and ignore completely
the spherical wind from the post-AGB star.
We also neglect heat conduction.
As will be evident from the results, we manage to show that CFW can also
explain the observed X-ray properties of some PNs. But before turning to compare
our results with observations in the coming sections, we describe the general flow
structure.
In these simulations we don't include ionizing radiation from the central star,
so the dense shell around the bubbles can cool to temperatures below $~10^4 \K$,
the typical temperature of visibly observed PN gas.

In reality, there are two opposite jets, one on each side of the equatorial plane.
We will present here only one quarter of the meridional plane. Namely,
we will present only one side of the equatorial plane, and one side of the symmetry axis.

\begin{figure}
\centering
\resizebox{0.45\textwidth}{!}{\includegraphics{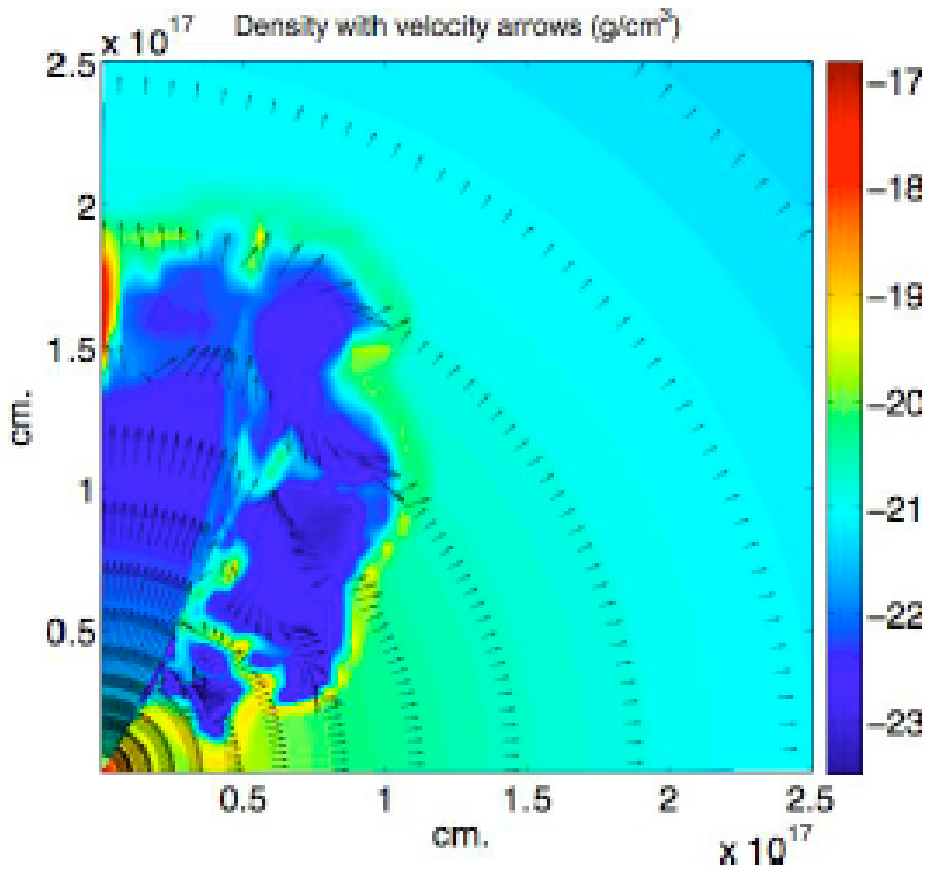}}
\centering
\resizebox{0.47\textwidth}{!}{\includegraphics{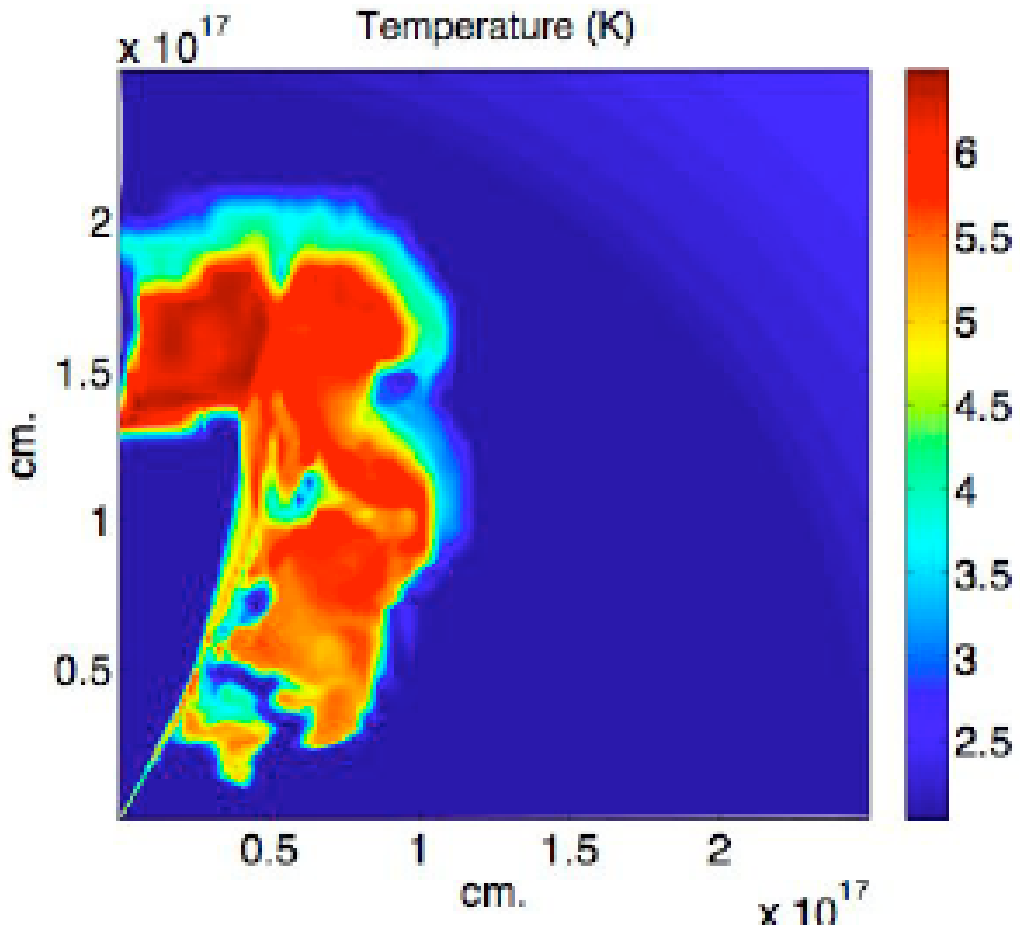}}
\centering
\resizebox{0.45\textwidth}{!}{\includegraphics{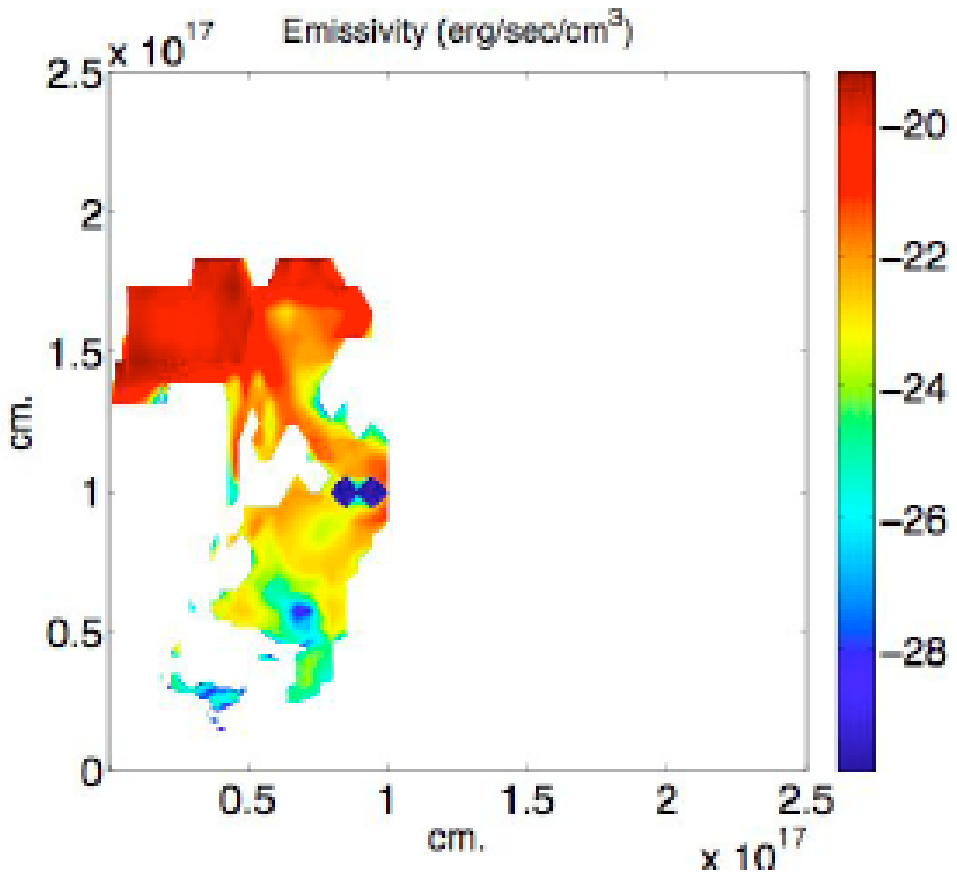}}
\centering
%
%
\caption{Density, temperature, and emissivity maps for
the run Y1 at $t= 1000 \yr$. Arrows indicate flow direction: $v > 200 \km \s^{-1}$
(long arrow), $ 20 < v \le 200 \km \s^{-1}$ (medium arrow), and $
v \le 20 \km \s^{-1}$ (short arrow).}
\label{Run1}
\end{figure}

In Fig. \ref{Run1} we plot the two dimensional maps of the density, the temperature, and the
emissivity (the X-ray power per unit volume in $\erg \s^{-1} \cm^{-3}$)
at $t = 1000 \yrs$ for run Y1 (see Table 1).
While the dense gas outside the hot bubble has a radiative cooling time much shorter
than the flow time, the hot interior of the bubbles has a long radiative cooling time.
The jets expand along the symmetry axis. The dense nebular gas in the equatorial
plane is not influenced much, and forms a narrow waist in the equatorial plane.
Such a structure is observed in many PNs.
As evident from Fig. \ref{Run1}, there is no X-ray emission on the jet's axis.
The X-ray comes mainly from the `cocoon' and the back-flow toward the center.
The cocoon$-$a well known feature of expanding jets$-$is the
slowly moving material around the expanding jet, which
is formed from the post-shock jet material and some ambient
matter (the slow-wind gas).
In our simulations the cocoon forms a low density large circulating flow to
the sides of the jets (in the axisymmetrical simulations the structure
is a torus).

\begin{figure}
\centering
\resizebox{0.48\textwidth}{!}{\includegraphics{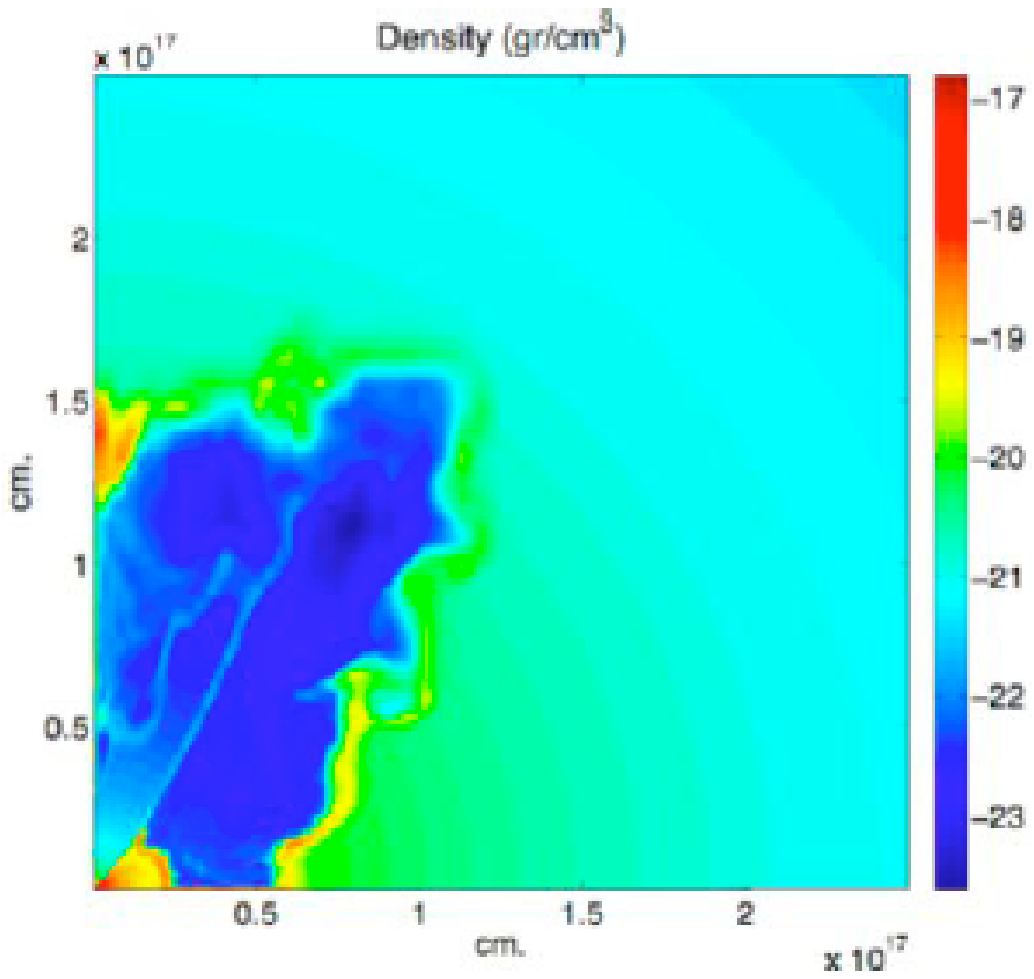}}
\centering
\resizebox{0.55\textwidth}{!}{\includegraphics{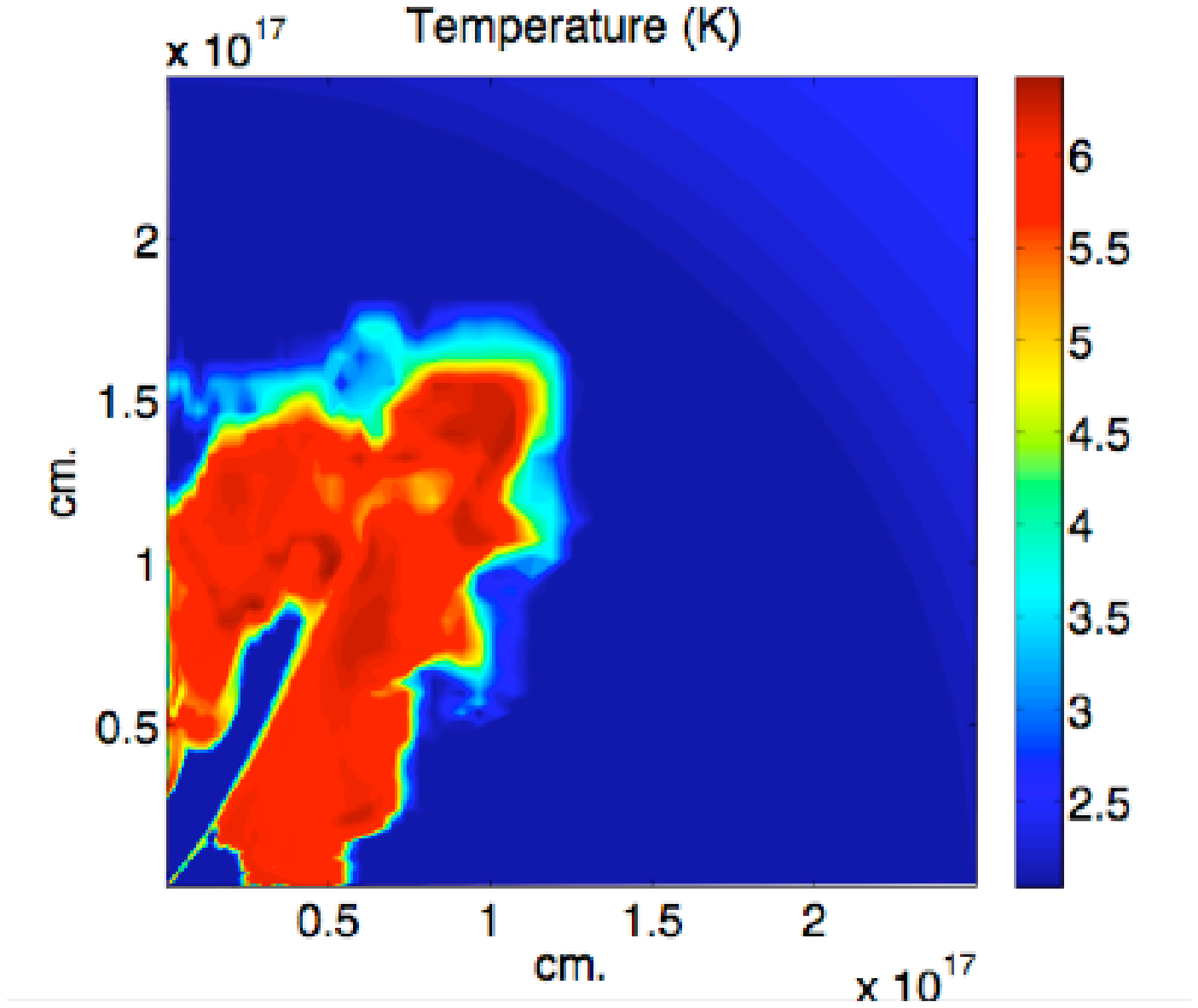}}
\centering
\resizebox{0.50\textwidth}{!}{\includegraphics{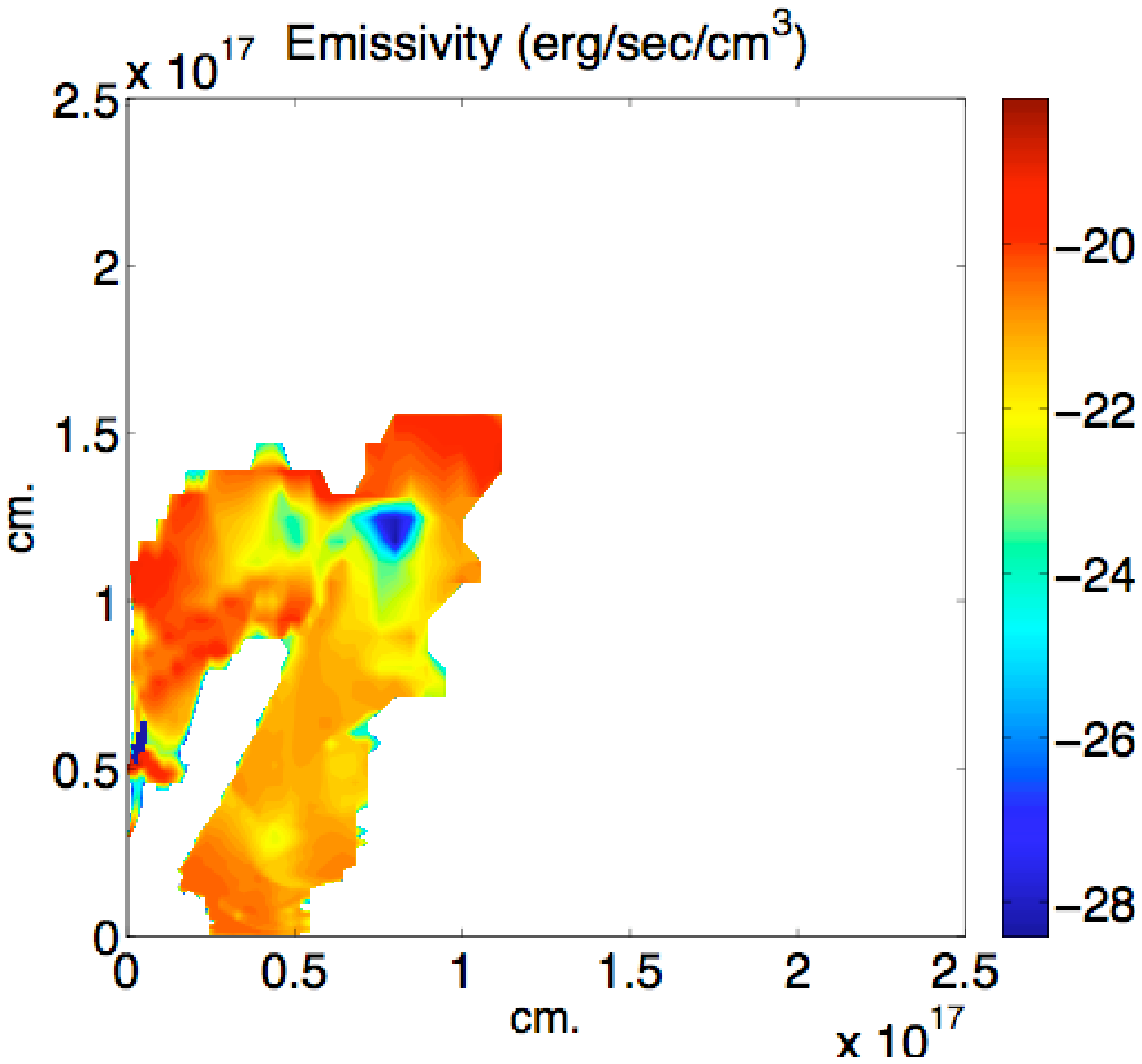}}
\caption{As Fig. \ref{Run1} but for run Y2.}
\label{Run2}
\end{figure}

\begin{figure}
\centering
\resizebox{0.51\textwidth}{!}{\includegraphics{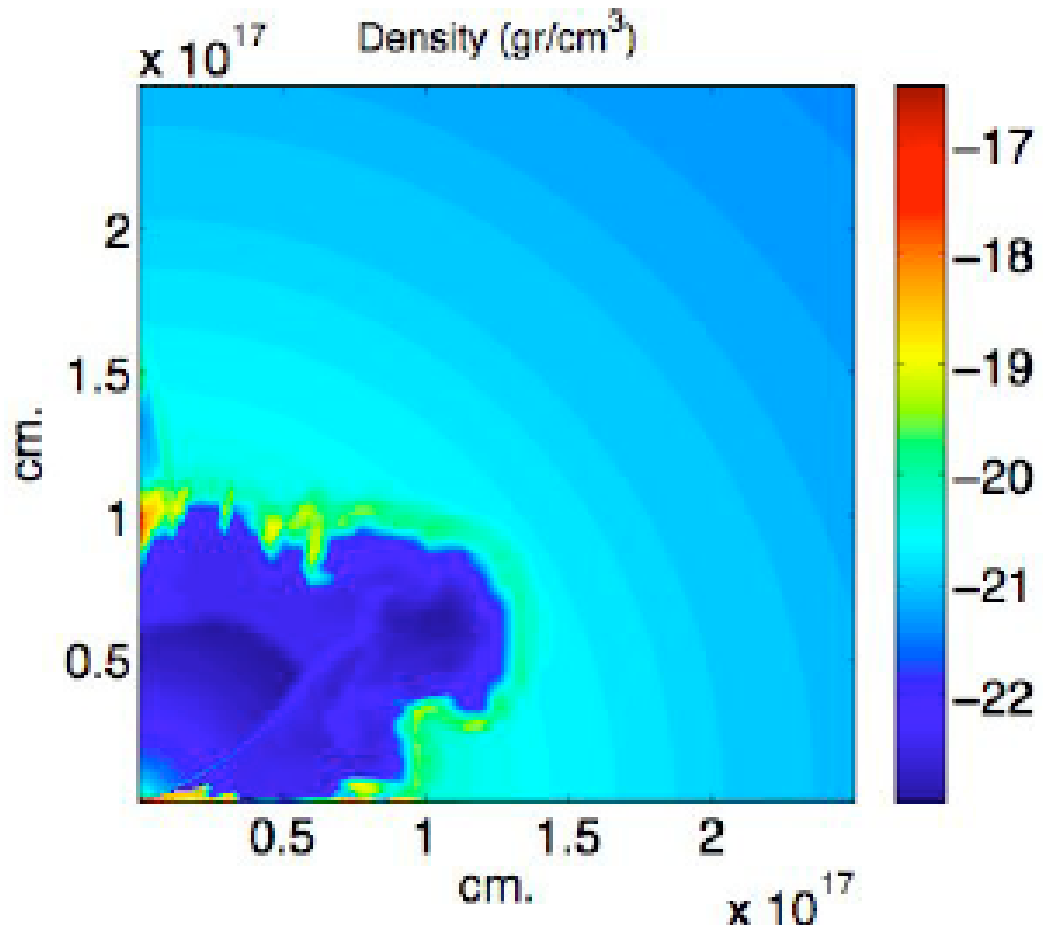}}
\centering
\resizebox{0.51\textwidth}{!}{\includegraphics{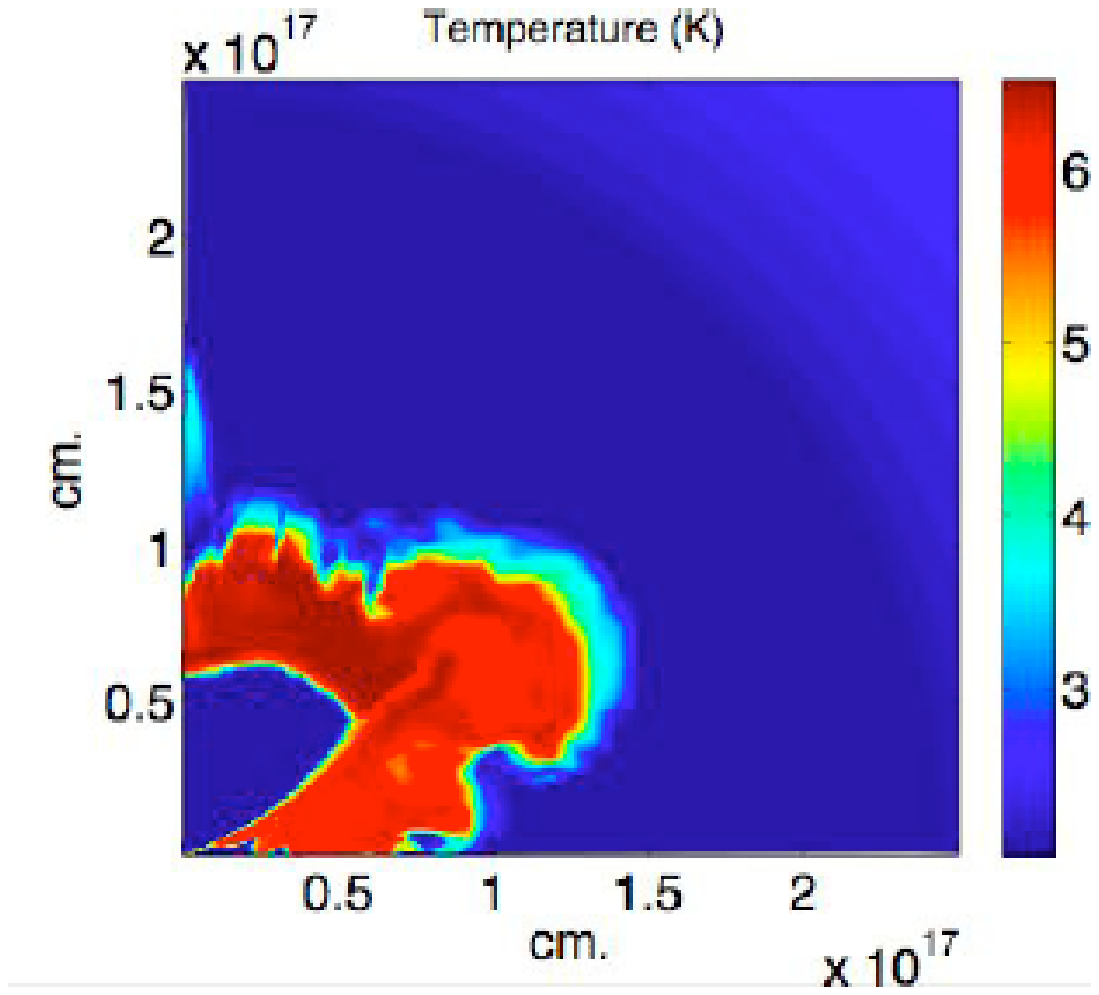}}
\centering
\resizebox{0.51\textwidth}{!}{\includegraphics{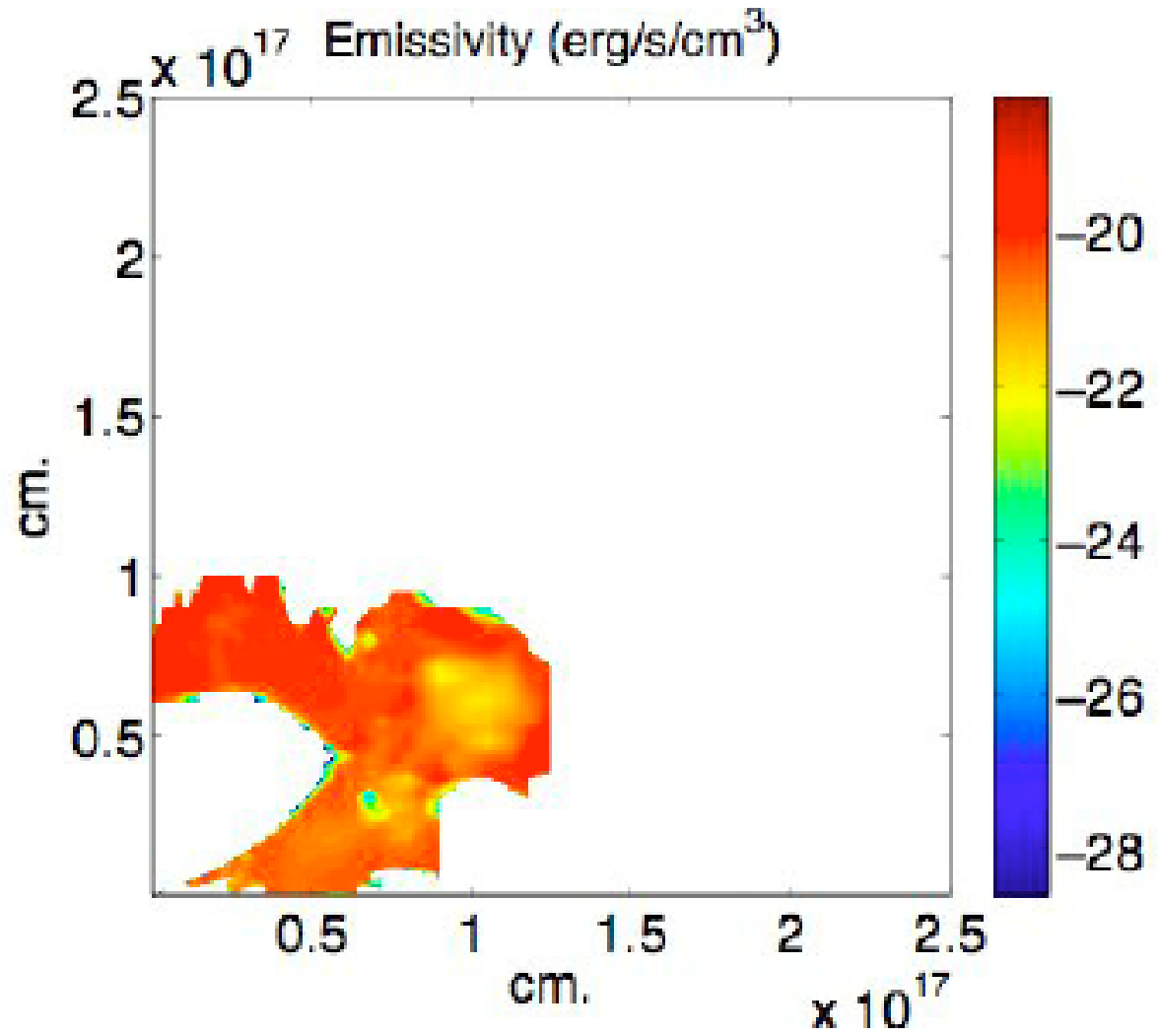}}
\caption{As Fig. \ref{Run1} but for run Y5.}
\label{Run5}
\end{figure}

The results of runs Y2 and Y5 are plotted on Fig. \ref{Run2} and Fig. \ref{Run5},
respectively.
As with run Y1, in these runs the jet is shut down gradually starting at
$t=\tau_s=1000 \yr$, and with $\tau_d = 1100 \yr$ in equation (\ref{power1}).
The difference between models Y1, Y2, and Y5 is the half opening
angle$-$Y1($\alpha = 30^\circ$), Y2($\alpha = 40^\circ$),
and Y5($\alpha = 75^\circ$).
In Fig. \ref{Run2} we see that X-ray emitting gas could be formed
on the symmetry axis.
We note that in some PNs, e.g., Menzel 3 (Kastner et al.\ 2003), X-ray emission is observed
along the symmetry axis. This might be because of a cocoon, as we find here, but more likely for
Menzel 3 it can also be emission inside the jet itself because of internal shocks.
In a future paper we will run denser jets and we will study emission within the jet itself.

In Fig. \ref{Run5} we see that X-ray emitting gas is not on the symmetry
axis. It comes mainly as in the Y1 model ($\alpha = 30^\circ$) e.g.
from the cocoon and the back-flow toward the center.
This is because the cocoon of such angles can build itself into this region.
We actually show that for very thin range of half opening angles ($30^\circ < \alpha < 50^\circ$),
the X-ray emitting gas could be formed on the jet's axis.

\begin{figure}
\centering
\resizebox{0.99\textwidth}{!}{\includegraphics{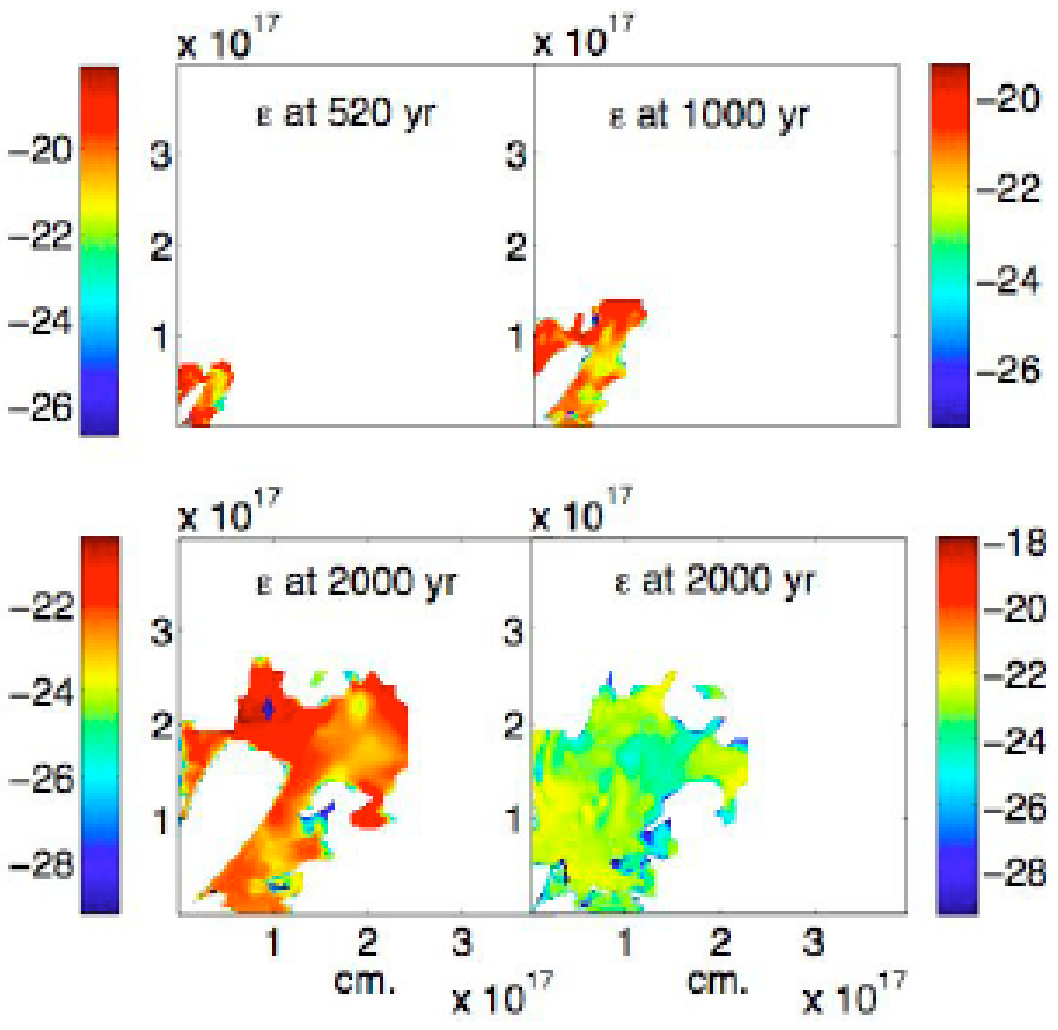}}
\caption{The emissivity ($\erg \s^{-1} \cm^{-3}$) evolution of run
Y8 and Y3 at three times. The lower left panel is run Y8 at
$t=2000 \yr$, while the lower right panel is run Y3 at $t=2000
\yr$. In run Y3 the wind is gradually shut down starting from
$t=1000 \yr$. In run Y8 it is constant at all times, but other
than that Y3 and Y8 are identical.}
\label{Fig4}
\end{figure}

In Fig. \ref{Fig4} we show the evolution of the emissivity of model
Y8 at three times, and compare the last frame with that of model Y3 at the same time.
Model Y8 is similar to run Y3, but the jet is not shut down.
In all simulations the temperature map is very similar to
the emissivity map because the hot gas is responsible for the X-ray emission.

\section{DEPENDANCE OF X-RAY PROPERTIES ON THE CFW PARAMETERS}

\subsection{Dependence on the opening angle}
\label{angle}

The X-ray luminosity is calculated by integrating the emissivity
over the hot regions in the PN in the energy range $0.2-10 \kev$,
but ignoring regions with $T<10^5 \K$ (Using the Formulae
in Akashi et al 2006, 2007). We choose this range to
reflect the sensitivity regime of the {\it Chandra}
and {\it XMM-Newton} telescopes.

\begin{figure}
\centering
\resizebox{0.89\textwidth}{!}{\includegraphics{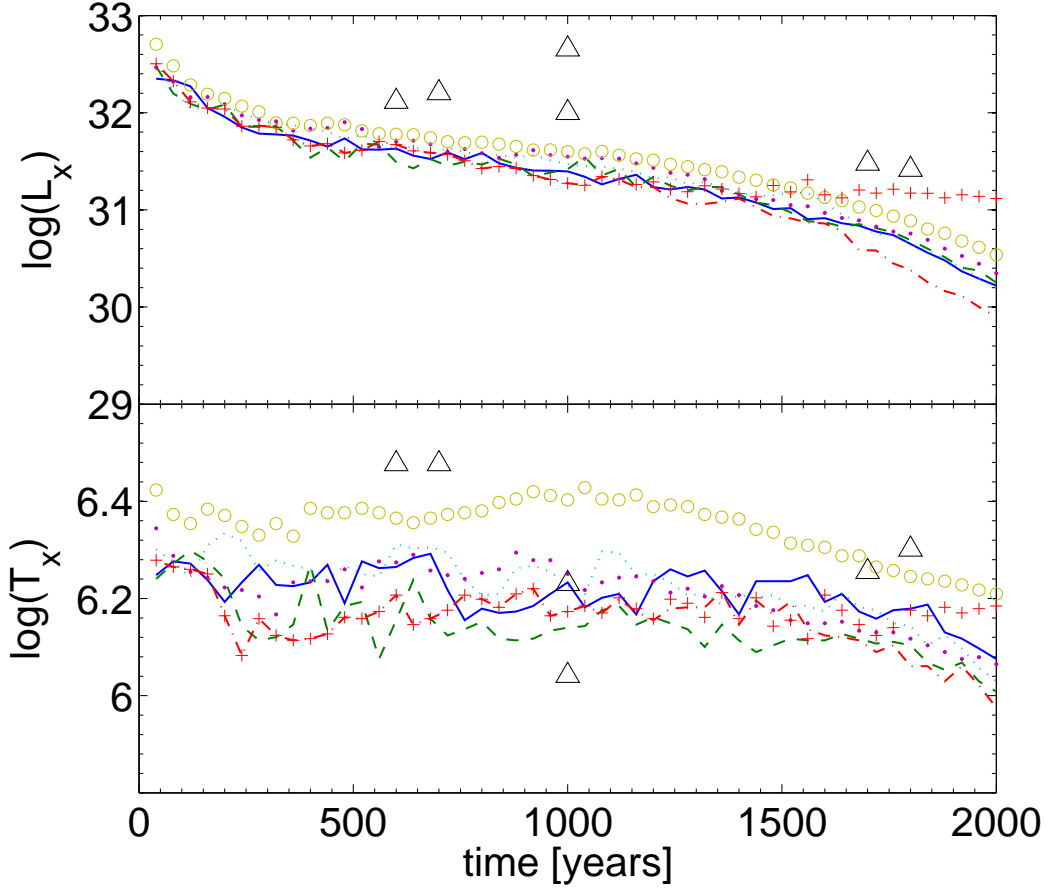}}
\caption{Top: X-ray luminosity as a function of time for PNs with different opening angle
from $30^\circ $ to $90^\circ $, as listed in Table 1:
models Y1 (solid), Y2 (dashed), Y3 (dashed-dotted), Y4 (dotted), Y5 (points), and Y6 (circles).
The plus signs are for the run Y8, which is Y3 but the jet is not shut down.
Bottom: X-ray temperature as a function of time for PNs with different opening angle
from $30^\circ $ to $90^\circ $, as listed in Table 1:
Y1 (solid), Y2 (dashed), Y3 (dashed-dotted), Y4 (dotted), Y5 (points), and Y6 (circles).
The higher temperature for the spherical run (circles) result from that one large bubble
is formed (instead of two when there are two jets), and it expands less, hence adiabatic
cooling is less efficient.
Observations from Table 2 are plotted by triangles.}
\label{Fig5}
\end{figure}

In Figs. \ref{Fig5} we plot the X-ray luminosity
and the temperature of the X-ray emitting gas, respectively, as a
function of time for six values of the half opening angle $\alpha$,
as indicated in the figure captions following Table 1.
Triangle mark the properties of PNs from observations, as summarized in Table 2.
Note that $\alpha=90^\circ$ corresponds to a spherical fast wind. As evident
from these figures the X-ray luminosity and temperature weakly
depend on the opening angle in the range $30^\circ \la \alpha \le 90^\circ$.
Only the spherical case ($\alpha=90^\circ$) is somewhat
different in this range.
As noted in the previous section,
different opening angles do lead to very different morphologies,
in the nebular gas (the gas at $\sim 10^4 \K$ that is bright in
the visible band), and in the X-ray. We also run cases for
$10^\circ$ and $20^\circ$, but don't show them. For runs with
narrow jets the general morphology is very different because the
jets do not form hot bubbles, but rather rapidly (after $\sim
500-800 \yrs$) expand to large distances. The X-ray luminosity is
lower than in the wider jets cases.
\begin{table}

Table 2: X-ray properties of Planetary Nebulae

\bigskip
\begin{tabular}{|l|c|c|c|c|}
\hline
\#&PN & $L_x$ & $T_x$ & Dynamical Age \\
 &  & $ 10^{32}$ erg s$^{-1}$ & $ 10^{6}$ K & yr\\
\hline
1& NGC 7027(PN G084.9-03.4) &1.3 &3 &600 \\
\hline
2& BD +30 3639(PN G064.7+05.0) &1.6 &3&700 \\
\hline
3&NGC 7026(PN G096.4+29.9) & 4.5 &1.1 & $<1000$  \\
\hline
4&NGC 6543(PN G096.4+29.9) &1.0&1.7 &1000  \\
\hline
5 &NGC 7009(PN G037.7-34.5) &0.3 &1.8 &1700\\
\hline
6 &NGC 2392 (PN G197.8+17.3) &0.26 &2 & 1800\\
\hline
7 &NGC 40 (PN G120.0+09.8) &0.024 &1.5 & 5000\\
\hline
\end{tabular}

\footnotesize
\bigskip
The parameters of the first three PNs and NGC 7009 are summarized
by Soker \& Kastner (2003). The data for NGC 2392 are from
Guerrero at al.\ (2005), for NGC 40 from Kastner et al.\ (2005),
and for NGC 7026 from Gruendl et al. (2006). \normalsize
\end{table}

The CFW total kinetic luminosity $\dot E_2 = 2 \times \frac{1}{2} \dot M_2 v_2^2$
(a factor 2 for the two jets),
is the same in all cases plotted on Fig. \ref{Fig5},
as the mass loss rate and jets' speed were held constant in these runs.
Despite the different large scale morphologies between the simulated cases,
and the instabilities that result in short-time variation of the X-ray luminosity,
the X-ray luminosity is similar in the runs that form bubbles
and have the same jet kinetic luminosity. Notice how the X-ray light
curve changes its behavior after $1000 \yrs$, as a result
of the decrease in the mass loss rate (eq. \ref{power1}).

\subsection{Dependence on the CFW mass loss rate}
\label{mdot}

\begin{figure}
\centering
\resizebox{0.89\textwidth}{!}{\includegraphics{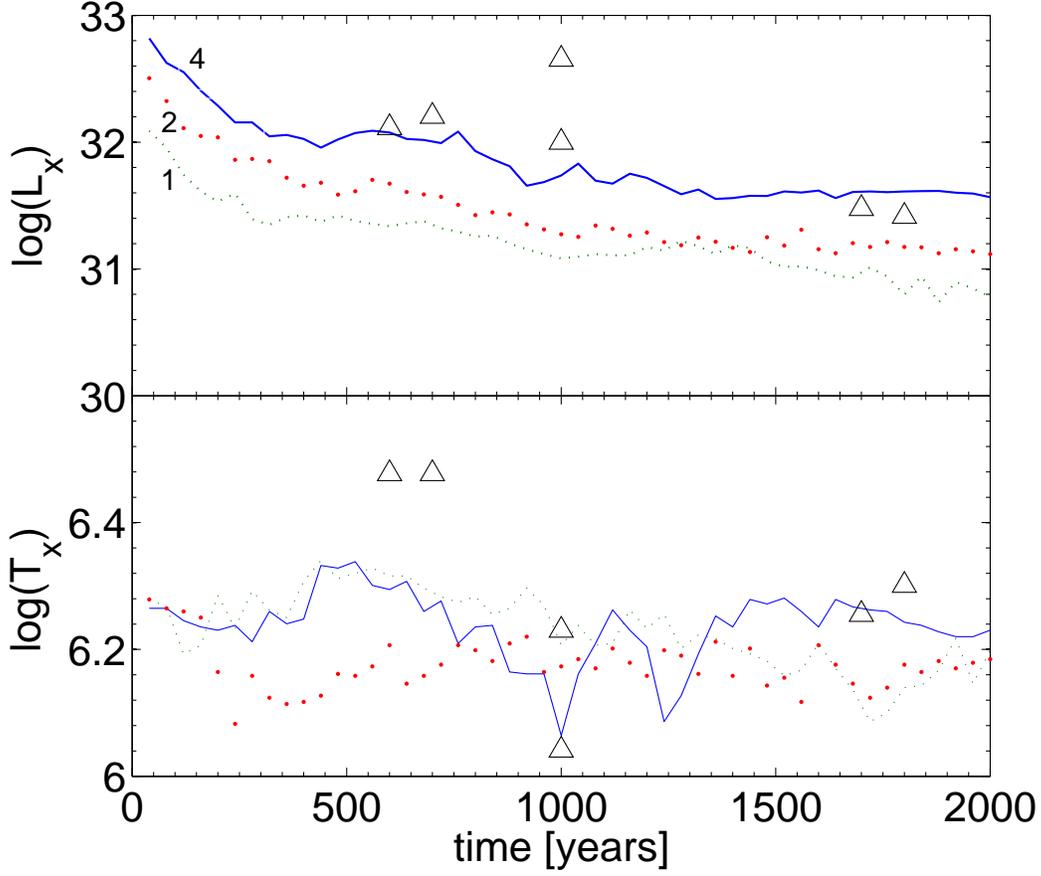}}

\caption{Top: X-ray luminosity as a function of time for PNs with different mass loss
rate of the CFW (jet): models Y7 (dotted), Y8 (=Y3) (points), Y9 (solid).
The numbers 1, 2, and 4 are the mass loss rates we take in units of $10^{-7} M_\odot \yr^{-1}$.
Bottom: X-ray temperature as a function of time for PNs with different mass loss
rate of the CFW (jet): models Y7 (dotted), Y8 (=Y3) (points), Y9 (solid).
In these runs the CFW was not shut down.
Observations from table 2 are plotted by triangles.}
\label{Fig6}
\end{figure}

In Figure \ref{Fig6} we present 3 simulations in which we take
$v_2=500 \km \s^{-1}$, $\alpha = 50^\circ$, $\dot M_1 = 1 \times 10^{-5} M_\odot \yr^{-1}$,
$v_1=10 \km \s^{-1}$, but we vary $\dot M_2$: $1 \times 10^{-7} M_\odot \yr^{-1}$,
$2 \times 10^{-7} M_\odot \yr^{-1}$, and $4 \times 10^{-7} M_\odot \yr^{-1}$,
Namely ,models Y7, Y8, and Y9.
We see that a CFW  with velocity of $500 \km \s^{-1}$ can fit observations if its
mass loss rate is up to $ (2-3) \times 10^{-7} M_\odot \yr^{-1}$.

\subsection{Dependence on the CFW velocity}
\label{speed}

\begin{figure}
\centering
\resizebox{0.89\textwidth}{!}{\includegraphics{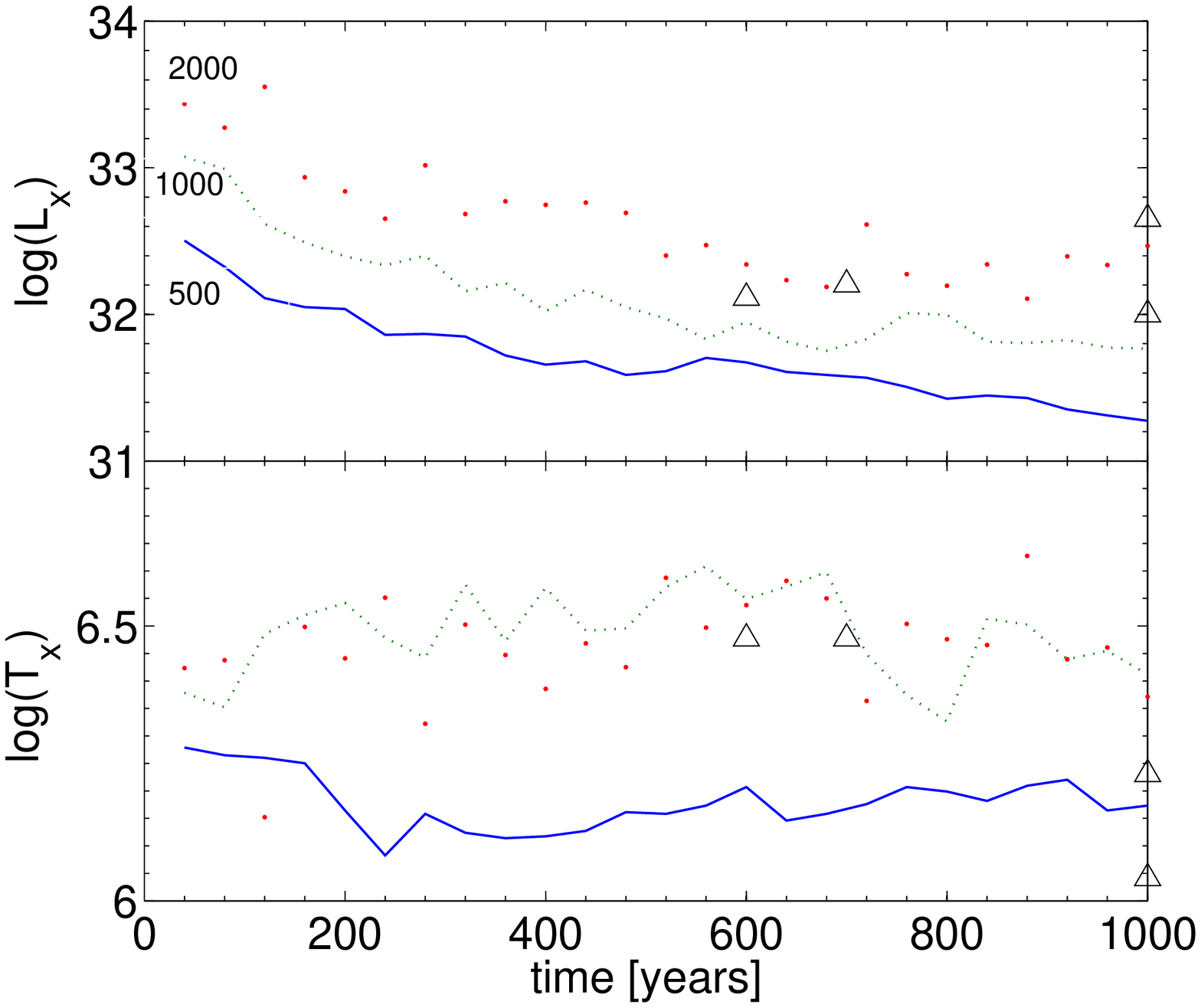}}
\caption{Top: The X-ray luminosity as a function of time for models with
different CFW (jet) velocities. The numbers shown in this panel are the
jet speeds in units $\km \s^{-1}$. Bottom: The X-ray temperature as a function
of time for the same models. Presented are models Y8 (solid), Y11 (dotted), Y12 (points).
Observations from table 2 are plotted by triangles.}
\label{Fig7}
\end{figure}

In Fig. \ref{Fig7} we present three simulations in which we vary the CFW (jet) speed.
We take $\dot M_2 = 2 \times 10^{-7} M_\odot \yr^{-1}$, $\alpha = 50^\circ$,
$\dot M_1 = 1 \times 10^{-5} M_\odot \yr^{-1}$, $v_1=10 \km \s^{-1}$,
and we vary $ v_2$: $500 \km \s^{-1}$, $1000 \km \s^{-1}$, $2000 \km \s^{-1}$.
Namely we show the models Y8, Y11, and Y12.
We find that a CFW with $<2 \times 10^{-7} M_\odot \yr^{-1}$ must be fast
$\sim 1500-2000 \km \s^{-1}$ in order to explain present observations.
In run Y12 the jet speed is four times as fast as in run Y8, which implies that the
dimidiate post-shock temperature is 16 times higher. However, in run Y12 the energy is
much higher and the bubble formed by the jet expands much faster, resulting in substantial
adiabatic cooling.
This is the reason for a moderate temperature differences, a factor of $\la 2$,
between the runs with large velocity differences.

\begin{figure}
\centering
\resizebox{0.89\textwidth}{!}{\includegraphics{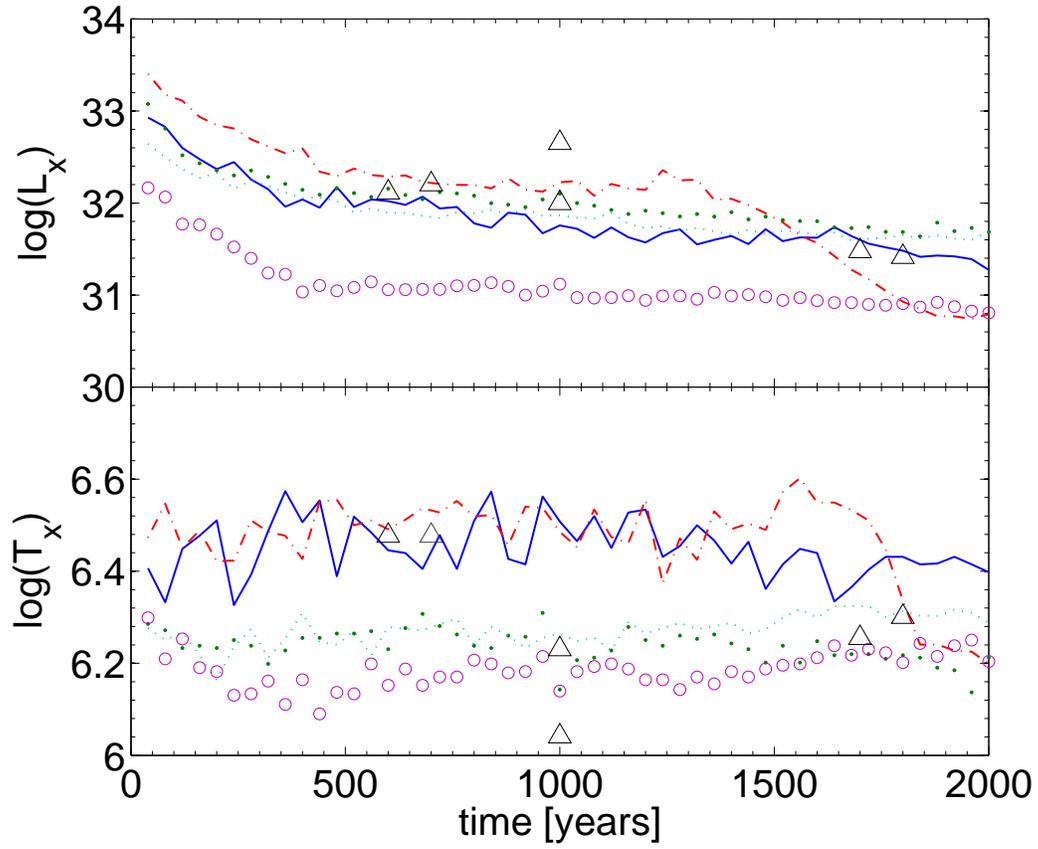}}
\caption{Top: X-ray luminosity as a function of time for PNs  for the models
Y13 (solid), Y14 (points), Y15 (dashed-dotted), Y16 (dotted), Y17 (circles).
Bottom: X-ray temperature as a function of time for the same models in the upper
panel. Observations from table 2 are plotted by triangles.}
\label{Fig8}
\end{figure}

To further explore the influence of the different parameters we present in
Fig. \ref{Fig8} the X-ray luminosity and temperature for models
Y13-Y17. From these we learn the following: (i)The slow wind must
be dense enough $\geq 10^{-5} M_\odot \yr^{-1}$ so the X-ray emitting
gas cannot escape from its envelope. (ii) Jet with ($300 \km \s^{-1} < v_2 < 900 \km \s^{-1}$)
fits well observations. (iii) Jet with ($10^{-7} M_\odot \yr^{-1}<\dot M_2< 8 \times 10^{-7} M_\odot \yr^{-1}$)
also fits well observations.

\section{DISCUSSION AND SUMMARY}

We summarize the open questions regarding the extended X-ray emission from
PNs and the implications of our results to these questions.

\subsection{ The source of the X-ray emitting gas.}
The source of the X-ray emitting gas can be the wind blown by the post-AGB
central star, or by the star during the PN phase (in both cases the wind is expected to be
spherical more or less),
or the source can be jets (or CFW for a collimated fast wind) blown by a
companion (Soker \& Kastner 2003).
Our results clearly show that jets blown by a companion at speeds of
$v_2 \simeq 500-3000 \km \s^{-1}$ can account for the properties of the extended X-ray emission.

Our estimate is that both post-AGB wind and jets occur in different PNs to different degree.
We predict that extended X-ray emission will be detected in some pre-PN systems
where the central star is too large (therefore the wind is too slow)
to eject the X-ray emitting gas.
The source of the X-ray emitting gas in these cases are jets blown
by a companion.
These pre-PN systems have a large extinction, and it is not easy to
detect the X-ray emission, but still possible.
We encourage an observational effort to detect X-ray emission from post-AGB stars.

\subsection{ The temperature of the X-ray emitting gas.}
If the source of the X-ray emitting gas are jets cooling by heat conduction
or mixing (Soker 1994; Zhekov \& Perinotto 1996; Steffen et al. 2005;
Chu et al. 1997) is not required.
Our results show that the postshock gas can cool adiabatically to the observed temperatures
as the bubbles expand.
Because adiabatic cooling is important, the temperature decreases with time.
This is compatible with the claimed of Kastner (2008) and Kastner et al. (2008).
This holds also if the source of the X-ray emitting gas is the wind blown by the central
star before the PN phase (Akashi et al. 2006, 2007).
Heat conduction is required only if the X-ray emitting gas is the fast
($\sim 2000 \km\s^{-1}$) wind blown by the central star during the PN phase.
Clearly heat conduction does not play a role if the stellar wind is slow.
This is the case with the strong X-ray emitting PN BD +30.3639 (PN G064.7+05.0), that
currently has a wind speed of $700 \km \s^{-1}$ (Leuenhagen et al. 1996),
much slower than required in the heat conduction model.

Our results predict that future observations will find PNs and pre-PNs with
X-ray emitting gas temperature higher than the post shock of the {\it present}
central wind speed. In these cases the source of the X-ray emitting gas must
be jets blown by the companion.
Again, observations of pre-PN objects are encourage.

In any case, our results do not rule out heat conduction and mixing.
Heat conduction might play some role in all these cases despite the presence of
magnetic field. It is hard to calculate the extend to which heat conduction
plays a role, because reconnection of magnetic field lines
between the cool nebular gas and the hot bubbles must take place to allow
heat conduction between the two media.

\subsection{ Morphology.}
The X-ray emitting gas has an asymmetrical morphology. In some
cases a pair of bubbles is observed, e.g., NGC 6543 (Chu et al.\ 2001).
It is possible that the bubbles were formed previous to the ejection of the X-ray
emitting gas, and the X-ray emitting gas just fill the bubbles.
This must be the case if the source of the X-ray emitting gas is the central stellar wind.
Our results suggest that it is possible that in some PNs the X-ray emitting gas itself formed
the bubbles. Namely, the shocked jets' material.

In a small number of cases we expect that both jets and the central stellar wind
contribute to the X-ray emission.
In these cases two-temperature gas will fit observations much better than a
single-temperature gas.
In addition, the two components will have different morphologies, with one showing
a more pronounced bipolar structure.

\acknowledgments
We thank John Blondin for his immense help with the numerical code.
This research was supported in part by the Asher Fund for Space Research at the
Technion.

\end{document}